# Molecular Dynamics Study of Liquid Condensation on Nano-structured Sinusoidal Hybrid Wetting Surfaces

Taskin Mehereen[1, a)], Shorup Chanda[1, b)], Afrina Ayrin Nitu[1, c)], Jubaer Tanjil Jami[1, d)], Rafia Rizwana Rahim[1,2, e)], Md Ashiqur Rahman[1, f)]

Author Affiliations

[1]*Department of Mechanical Engineering, Bangladesh University of Engineering and technology, Dhaka-1000, Bangladesh*
[2]*Oden Institute for Computational Engineering and Sciences, The University of Texas at Austin, Austin, TX-78712*

Author Emails

a) taskin2560@gmail.com
b) scshorup.chanda@gmail.com
c) afrinanitu71@gmail.com
d) jamijubaer@gmail.com
e) rafia.rizwana.rahim@utexas.edu
f) Corresponding author: ashiqurrahman@me.buet.ac.bd

**Abstract.** Although real surfaces exhibit intricate topologies at the nanoscale, rough surface consideration is often overlooked in nanoscale heat transfer studies. Superimposed sinusoidal functions effectively model the complexity of these surfaces. This study investigates the impact of sinusoidal roughness on liquid argon condensation over a functional gradient wetting (FGW) surface with 84% hydrophilic content using molecular dynamics simulations. Argon atoms are confined between two platinum substrates: a flat lower substrate heated to 130K and a rough upper substrate at 90K. Key metrics of the nanoscale condensation process, such as nucleation, surface heat flux, and total energy per atom, are analyzed. Rough surfaces significantly enhance nucleation, nearly doubling cluster counts compared to smooth surfaces and achieving a more extended atomic density profile with a peak of approximately 0.07 atoms/nm$^3$ and improved heat flux (450 MW/m$^2$). Stronger atom-surface interactions also lead to more efficient energy dissipation ($-0.06$ eV per atom). These findings underscore the importance of surface roughness in optimizing condensation and heat transfer, offering a more accurate representation of surface textures and a basis for designing surfaces that achieve superior heat transfer performance.

## INTRODUCTION

Efficient thermal energy management is crucial across industries, from power plants and refrigeration systems to microelectronic cooling and nanofluidic biochips [1–4]. As heat dissipation becomes a major obstacle, particularly in nano-scale devices, single-phase cooling methods are often inadequate [5] Latent heat transfer, such as condensation, offers a more effective solution by converting vapor into liquid, which is key for phase-change cooling systems [6,7]. Condensation is commonly utilized in devices like heat pipes for cooling CPUs, where heat transfers from vapor to a liquid coolant via a conductive surface, creating a temperature gradient [8].

Research on nanoscale condensation showed that surface modifications, particularly in wettability and design, can optimize heat transfer. Condensation occurs through two main mechanisms: filmwise condensation (FWC) on hydrophilic surfaces, and dropwise condensation (DWC) on hydrophobic surfaces. Hybrid wetting surfaces that combine these properties can maximize heat transfer [9–13]. Studies using hybrid surfaces, such as patterned and Functionally Gradient Wettability (FGW) surfaces, revealed that stronger hydrophilicity combined with weaker hydrophobicity yielded optimal results [14]. Surface morphology, such as the addition of nanostructures, also enhanced heat transfer, as shown in condensation and boiling simulations [15,16]. Nanostructures enhance heat transfer by increasing surface area and interaction between the surface and the fluid [17,18].

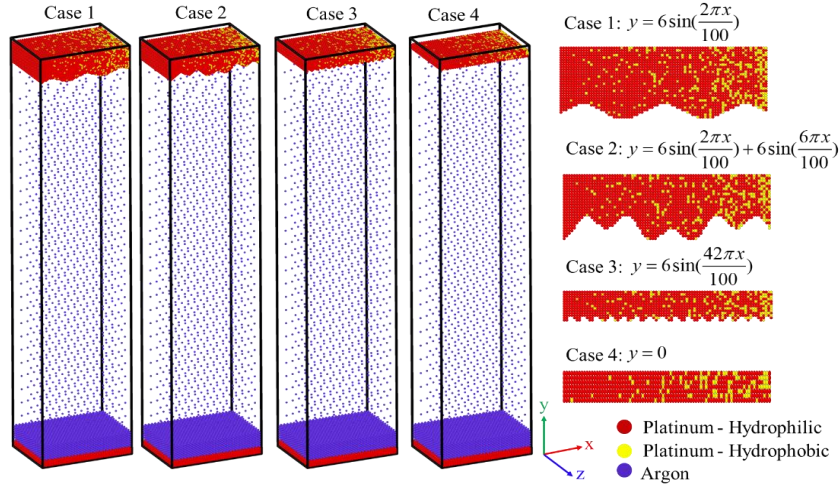

**FIGURE 1.** Graphical representation of the simulation model.

Although much research focuses on simple nanostructures, more complex surface geometries, such as those modeled using superimposed sinusoidal functions, have been shown to improve convection heat transfer [18]. This approach could also be applied to condensation studies. The Fourier transform can decompose complex surface geometries into sine waves for easier analysis, and the inverse Fourier transform can be used to reconstruct optimized surfaces for enhanced heat transfer. While this method has been used in convection studies in the nanoscale [18], it has not been widely explored in condensation. Thus, this work addresses the gap by proposing a molecular dynamics study of Ar atoms condensing on a series of rough surfaces having topological configurations of superimposed sinusoidal functions. The condensing surface has an FGW–governed hydrophilic-hydrophobic composition. The work compares the effect of the order of the sine functions in enhancing the condensation heat transfer.

## METHODOLOGY AND VALIDATION

The simulation setup depicted in Fig. 1 features two solid platinum (Pt) substrates and a pure argon (Ar) fluid sample. The Pt substrates, arranged in a face-centered cubic (FCC) (100) lattice with a density of 21,450 kg/m³, are positioned at the top and bottom of the simulation domain, which measures 16 nm × 101 nm × 8 nm ($x \times y \times z$). The bottom Pt substrate holds a 4 nm-thick layer of liquid argon which has a density 1378 kg/m³ at temperature of 90 K, while the remaining space is filled with saturated argon vapor, with a density value 6.751 kg/m³ at 90 K. To maintain stability, the atoms in the bottommost layer of the Pt substrate are fixed, preventing them from leaving the domain, while the next two layers serve as a heat source. The subsequent five layers are responsible for conducting heat to the argon. The top substrate follows the same configuration as the bottom one. Initially, the system comprises 26586 Pt atoms, 10395 liquid Ar atoms, and 1336 vapor Ar atoms, all arranged in FCC crystal structures based on their respective densities.

The simulation employs the 12-6 Lennard-Jones (L-J) potential, as demonstrated by Paul *et al.* [14], to calculate the intermolecular interactions. The system is constructed using an input script for the Large-scale Atomic/Molecular Massively Parallel Simulator (LAMMPS) [19]. The setup is modeled as an L-J system with periodic boundaries along the $x$ and $z$ directions and fixed boundaries along the $y$ direction, ensuring the stability of the simulation domain. In the L-J potential, the intermolecular potential depends on parameters such as $\varepsilon_{ij}$, which represents the depth of the potential well, $\sigma_{ij}$, which defines the distance where the interparticle potential equals zero, and $r_{ij}$, which denotes the distance between two atoms $i$ and $j$. For this simulation, a cutoff distance ($r_{cut}$) of $4.5\sigma_{Ar\text{-}Ar}$ is used to enhance computational performance, as suggested by [14].

**TABLE 1.** L-J model parameters for Ar and Pt.

| Atom | $\varepsilon$ (eV) | $\sigma$ (nm) |
|---|---|---|
| Ar | 0.0104 | 0.3405 |
| Pt | 0.5211 | 0.2475 |

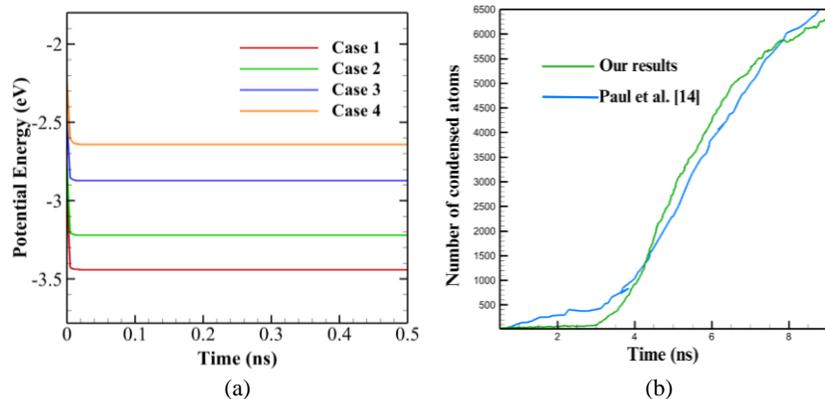

(a) (b)
**FIGURE 2.** (a) Variation of system potential energy during equilibration simulation, (b) Comparison between the reported values in the literature and those from our simulation for condensed atoms of liquid argon

The literature values [20] for the length parameters and potential well depths are summarized in Table 1. Following the method by Hens *et al.* [21], the altered magnitudes of the potential well depth for hydrophilic and hydrophobic Pt atoms interacting with argon atoms are included. Specifically, the $\varepsilon_{Ar-Pt}$ is 0.0208 eV for hydrophilic platinum and 0.0052 eV for hydrophobic platinum. In this context, the depth of the potential well ($\varepsilon$) indicates the interaction strength between two atoms. Argon, with a lower $\varepsilon$ value of 0.0104 eV, exhibits weaker interatomic interactions compared to platinum, which has a much higher $\varepsilon$ value of 0.5211 eV. The $\sigma$ parameter, which reflects the atomic sizes and characteristic distances between atoms, is 0.3405 nm for argon and 0.2475 nm for platinum.

To model the effect of surface roughness on different configurations of hybrid wetting condensing surfaces, sinusoidal functions are used to define the surface structure, maintaining a fixed 84% hydrophilic surface coverage. In alignment with the methodology outlined by Paul *et al.* [14], the entire solid substrate at the bottom within the specified dimensions is substituted with hydrophilic atoms. Fig. 1 demonstrates different cases where sinusoidal functions are used to model the roughness patterns. These cases vary based on the frequency and amplitude of the sinusoidal functions applied, representing different levels of surface roughness. Each case presents a specific configuration of roughness, which is mathematically defined as a series of sinusoidal waves cut along the *x*- and *y*-axes. Case 1 shows a lower-frequency sinusoidal function, providing a surface with moderate roughness. In Case 2, a more complex roughness pattern is introduced by superimposing an additional term to the sinusoidal function with a higher frequency, increasing the number of surface features. Case 3 presents a higher-frequency uniform sinusoidal function, creating a finer roughness pattern with more condensed surface features. Case 4 represents the smooth surface used as a control, where no sinusoidal roughness is applied. The sinusoidal functions are mathematically defined to represent nanoscale surface structures while maintaining a consistent wetting characteristic of 84% hydrophilic platinum atoms across all surfaces. The input scripts for the FGW model are generated using the methods in [14], which enables precise manipulation of surface properties, allowing for customized wetting characteristics, such as the percentage of hydrophilic and hydrophobic regions, as required.

The simulation process in this study is divided into two stages: equilibrium molecular dynamics (EMD) and non-equilibrium molecular dynamics (NEMD). In the EMD stage, the system is run using the canonical ensemble for 0.5 ns, with the temperature maintained at 90 K by a Langevin thermostat. Fig. 2(a) illustrates the trend in potential energy over time during equilibration for different cases. As the system equilibrates, the potential energy decreases gradually before stabilizing, a behavior observed consistently across all four cases. After achieving thermal equilibrium, NEMD session begins, conducted under the microcanonical ensemble for 2.0 ns. In this phase, the temperatures of the phantom atoms in lower and upper substrates are set to 130 K and 90 K, respectively, controlled by a Langevin thermostat. The boiling temperature of liquid argon is represented by 130K; at 90K, the argon is in a liquid state when it condenses onto the top substrate. During the NEMD process, thermal energy is transferred from the lower substrate to the liquid atoms, leading to evaporation. These atoms then release energy to the upper substrate, undergoing condensation. The lower substrate emulates the evaporating surface, while the upper substrate emulates the condensing surface. To ensure the condensing surface's performance remains unaffected by the evaporation process, the evaporating surface wetting behavior is kept constant across all cases. The NEMD session lasts for 6.6 ns, with the positions and velocities of atoms determined using Newton's equations of motion. A velocity-verlet algorithm, with a timestep of 5 fs, is applied to solve the equations of atomic motion. Visualization of atomic trajectories is generated using the Open Visualization Tool (OVITO) [22] software. The condensation characteristics of various nanostructured surfaces in this study are evaluated using metrics commonly employed in molecular dynamics simulations, including system energy, nucleation, coalescence, condensate growth, surface heat flux, temperature distribution, condensate mass flux, and radial distribution function. To validate the simulation accuracy, the time history of the condensation number is compared with that of Paul *et al*. [14], confirming the close agreement and thereby validating the proposed methodology, shown in Fig. 2 (b).

## RESULTS AND DISCUSSION

The temporal evolution of the total energy of argon atoms, as shown in Fig. 3(a), reveals the impact of surface roughness on condensation. Across all cases, the total energy initially decreases during equilibration, stabilizing around −0.07 eV per atom by 0.5 ns. The condensation process triggers energy spikes due to droplet nucleation. For cases 1 to 3 (moderate to high roughness), the energy rises to −0.05 eV per atom at 0.85 ns and stabilizes at −0.07 eV per atom by 1.75 ns. Case 4 (smooth surface) shows a more significant energy rise, peaking at −0.025 eV per atom and stabilizing at −0.05 eV per atom. This suggests that surface roughness enhances system stability by facilitating frequent interactions between argon atoms and the surface, leading to greater energy dissipation. Rougher surfaces exhibit more negative total energy values, indicating stronger interactions and enhanced stability, while smoother surfaces reflect less negative energy values and reduced stability. Figures 3(b) and 4 illustrate how surface roughness influences nucleation, with rougher surfaces (e.g., Case 3) exhibiting more initial condensation clusters. The hydrophilic platinum regions (red) attract condensate atoms, creating initial clusters that act as seeding points for further condensation. In cases with greater roughness, more nucleation sites are available due to the increased surface area. Case 3, with high-frequency roughness, exhibits a rapid increase in clusters during the early stages of condensation.

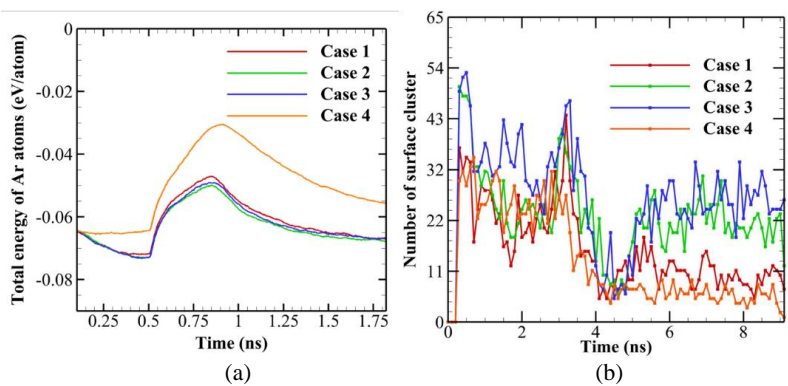

**FIGURE 3.** Time history of (a) total system energy, and (b) surface cluster for the various cases

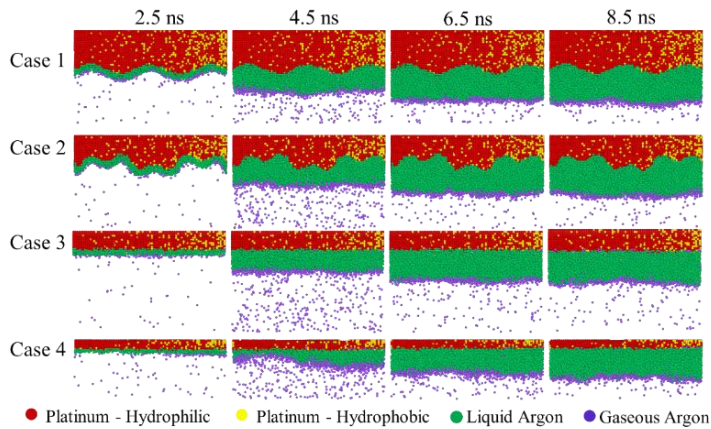

**FIGURE 4.** Atomic snapshots of the growth of condensate on the surfaces for the various cases

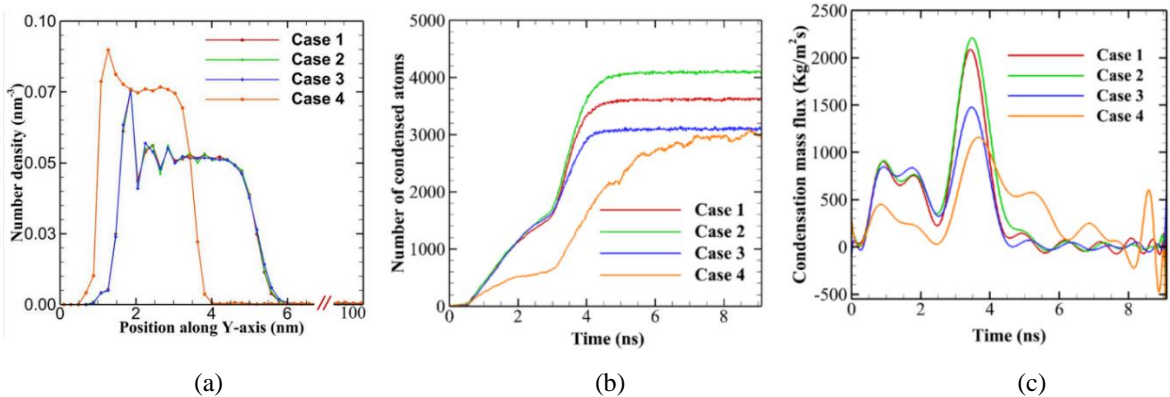

| (a) | (b) | (c) |

**FIGURE 5.** (a) Number density of argon atoms along the y-axis for all cases (b) Time history of condensation of argon atoms (b) Time history of condensation mass flux

Over time, these clusters coalesce and grow, resulting in larger but fewer droplets as condensation progresses. As shown in Fig. 3(b), rough surfaces (e.g., Case 3) maintain higher initial cluster counts, peaking around 54 clusters, which coalesce and reduce to approximately 22 to 25 clusters. Smoother surfaces (e.g., Case 4) show lower initial cluster counts, peaking around 32, with slower nucleation and fewer clusters (5 to 10). The behavior in rougher surfaces leads to more rapid nucleation, faster cluster coalescence, and larger droplets, while smoother surfaces exhibit slower growth and less dynamic clustering behavior.

Figure 5(a) illustrates the number density profile of argon atoms along the y-axis, highlighting the influence of surface roughness on the condensation process. For rough surfaces (Cases 1, 2, and 3), the atom density near the surface is lower, with a peak density of approximately 0.07 atoms/nm³. In contrast, the smoother surface (Case 4) exhibits a more gradual increase in atom density, reaching a peak of 0.09 atoms/nm³. Notably, for the rough surfaces, the number density remains elevated up to around 6 nm from the surface, suggesting an extended region of condensed atoms. This behavior is consistent with theoretical predictions that rough surfaces offer more nucleation sites, thereby enhancing condensation efficiency and promoting the formation of a condensed layer that extends farther from the surface. On the other hand, while the smoother surface shows a higher peak in atom density, the condensation is limited to a narrower region, with significant density observed only up to around 4 nm. These results indicate that smoother surfaces may lead to greater accumulation of atoms in close proximity to the surface, while rough surfaces facilitate condensation over a wider region. This distinction underscores the trade-off between surface roughness and the spatial distribution of the condensed phase, with rough surfaces supporting a broader, more extended condensation layer, whereas smooth surfaces concentrate condensed atoms in a more confined area.

The condensation performance is tracked by analyzing the number of condensed atoms and mass flux at each timestep, as illustrated in Fig. 5(b) and (c). Case 2, with the higher roughness, exhibits the greatest number of condensed atoms, demonstrating that surface roughness enhances condensation performance. Lower roughness (Case 1) results in higher condensation rates compared to the smoother surface (Case 4). The performance follows this descending order: Case 2, Case 1, Case 3, and Case 4. As roughness increases in Case 3, the surface begins to approximate the flat surface, causing the final condensation value to converge with that of Case 4. The condensation mass flux, calculated from the instantaneous slope of the curves in Fig. 5(b), and illustrated in Fig. 5(c), follows a similar trend. Case 2 shows the highest mass flux, indicating superior heat transfer performance, with the same descending performance order as Case 2, Case 1, Case 3, and Case 4. This confirms that surface roughness plays a critical role in enhancing condensation rates and heat transfer efficiency, at the hydrophilic fraction of 84%. Rougher surfaces consistently exhibit better condensation characteristics, emphasizing the importance of optimizing surface roughness in such processes.

The radial distribution function (RDF) for all the elements in all four cases is shown in Fig. 6. In this plot, the vertical axis represents the RDF, which indicates the probability of finding two atoms at a certain separation distance. The horizontal axis refers to the radial distance from a reference atom, measuring how the local structure of atoms changes as one moves outward from this atom. Case 2 has the highest number of condensed atoms (~4600), indicating the most effective condensation process. This aligns with the RDF in Case 2, where strong Ar-Pt interactions and clear Ar-Ar liquid peaks suggest significant condensation and atomic clustering. Case 1 comes second, with a plateau of around 4000 condensed atoms. This suggests good condensation behavior, though slightly less than Case

2. The RDF from Case 1 supports this, as it shows moderate Ar-Pt and Ar-Ar liquid interactions. Case 3 has fewer condensed atoms (~3700) and a slower rate of condensation. This matches the RDF for Case 3, which showed weaker Ar-Pt interactions, indicating that the surface roughness is not as conducive for condensation. Case 4 exhibits the lowest number of condensed atoms (~2500) and a notably slow condensation rate. Despite the strong Ar-Pt interactions shown in the RDF for Case 4, which would suggest better condensation, the number of condensed atoms is surprisingly lower. This suggests that while the flat surface allows for strong interactions at the surface level, it may not provide as many nucleation sites as the sinusoidal rough surfaces. In rough surfaces, valleys and ridges offer additional surface area and locations for condensation, which enhances the overall number of condensed atoms. The flat surface does not provide these advantages, limiting large-scale condensation despite strong local interactions. The strong Ar-Pt peaks in the RDF for Case 4 likely reflect the formation of a thin, dense layer of condensed argon atoms on the flat surface. However, the flat geometry does not promote further growth of condensed layers beyond this initial layer, leading to a plateau in the number of condensed atoms. The rough surfaces in the other cases, by contrast, allow for more extensive condensation due to their complex topography. If the goal is to maximize total condensation, rough surfaces (especially Case 2, which has the highest number of condensed atoms) are better. They create more nucleation sites, leading to more extensive condensation. If the goal is uniformity, tightly packed layers of condensed material, the flat surface (Case 4) might be better due to the stronger, more localized interactions between the substrate and condensed atoms, even though the total number of condensed atoms is lower.

Temperature profiles along various $y$-coordinates, shown in Fig. 7(a) reveal fluctuations between 90 K and 130 K, reflecting the dynamic molecular motion during phase transitions. These fluctuations result from localized heating or cooling as molecules absorb or release latent heat during condensation and evaporation. The Van der Waals interactions at phase boundaries play a key role in these temperature variations. Case 2 exhibits the smallest temperature fluctuations, indicating efficient heat dissipation due to its rough surface. In contrast, Case 4 shows the largest fluctuations, corresponding to less efficient heat transfer. Regions of intense condensation show peaks in temperature due to molecular collisions and increased kinetic energy, while regions of slower molecular motion, such as during clustering, show lower temperatures. The temperature differences between gas and liquid phases create distinct temperature gradients, further emphasizing the impact of surface roughness on phase change processes and thermal behavior. The heat flux $q_s$ across the condensing wall quantifies the rate at which heat is extracted during the condensation process. Fig. 7(b) shows that the maximum heat flux values follow the following in descending order as Case 2, Case 1, Case 3, and Case 4. The results align with the mass flux trends, highlighting that surface roughness significantly influences heat transfer performance. Despite maintaining an equal hydrophilic fraction across all cases, rougher surfaces demonstrate superior heat transfer capabilities. Case 2, with its high roughness, achieves the highest maximum condensing wall heat flux, while Case 4, with its smooth surface, exhibits the lowest. This trend remains consistent across all cases, with the differences in average heat flux values attributed primarily to surface roughness. These findings confirm that optimizing surface roughness is crucial for improving heat transfer efficiency.

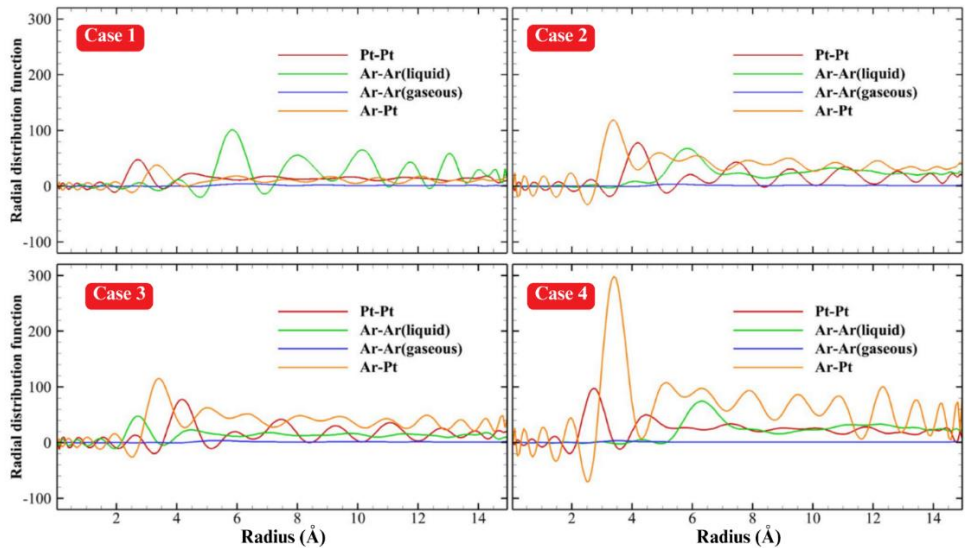

**FIGURE 6.** Radial Distribution Function with respect to Radius (Å) for Pt-Pt, Ar-Ar (Liquid and Gaseous), and Ar-Pt interactions across four cases with varying surface geometries

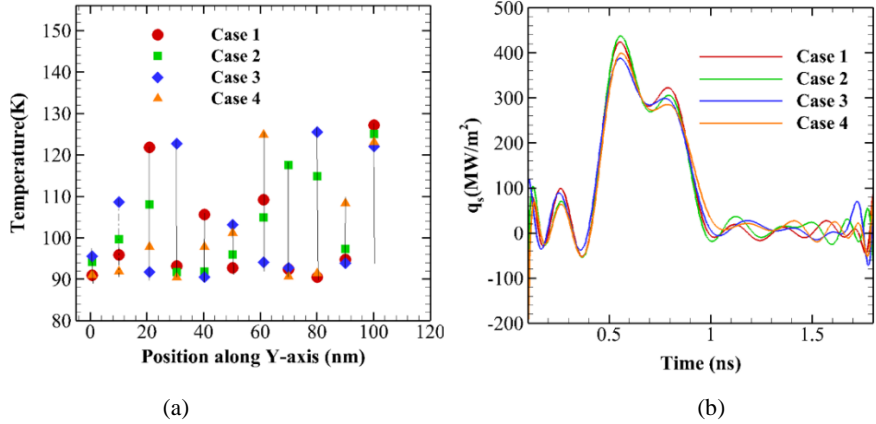

(a)                         (b)

**FIGURE 7.** (a) Temperature profile of the system along the direction of evaporation and condensation ($y$-axis) (b) Time history of surface heat flux $q_s$ across the condensing wall during the condensation process for all four cases

## CONCLUSIONS

In this study, the condensation characteristics of nanostructured surfaces with varying roughness are examined, focusing on their effects on nucleation, energy dissipation, and heat transfer performance during the condensation of argon atoms. Molecular dynamics simulations are employed to analyze key metrics such as nucleation rates, surface cluster growth, and energy profiles under different surface conditions. The results provide valuable insights into the relationship between surface morphology and condensation behavior, revealing both qualitative and quantitative impacts of surface roughness on thermal performance. The key conclusions drawn from this study are summarized as follows:

1. Rough surfaces exhibit significantly enhanced nucleation by providing more condensation sites for argon atoms, resulting in a more extended region of condensed atoms. These surfaces show a peak atomic density of approximately 0.07 atoms/nm³ near the surface, while the smooth surface reaches a higher peak density of around 0.09 atoms/nm³. However, the rougher surfaces facilitate condensation over a larger distance, extending up to about 6 nm from the surface, compared to only 4 nm for the smoother surface. This demonstrates that while smoother surfaces concentrate condensed atoms in a smaller region, rougher surfaces promote more widespread condensation, offering more nucleation sites that enhance condensation efficiency.
2. Surface roughness also directly impacts heat flux, as rougher surfaces facilitate better heat transfer during condensation. Case 2 (high-frequency sinusoidal surface) of the present study exhibited the highest condensing wall heat flux, aligning with its superior condensation mass flux, which peaked at 2200 kg/m²s, indicating efficient heat dissipation during phase change. The rougher surfaces increase surface interactions, leading to more efficient latent heat transfer and improved overall thermal performance.
3. The total energy behavior of the argon atoms during condensation reveals that rough surfaces promote stronger atom-surface interactions, resulting in more negative total energy values. It indicates that increased surface roughness facilitates more frequent and complex atom-surface interactions, improving system stability and condensation efficiency.

The findings from this study demonstrate that surface roughness plays a pivotal role in determining the efficiency of condensation processes. Optimizing surface roughness, therefore, emerges as a critical factor for improving condensation behavior and heat transfer in practical phase change processes.

## ACKNOWLEDGMENTS

The authors would like to express their sincere gratitude to the Department of Mechanical Engineering at the Bangladesh University of Engineering and Technology (BUET) for providing the resources and support necessary to conduct this research. Special thanks go to the members of the research group for their valuable insights and collaboration.

# REFERENCES

1. J.M. Beér, "High efficiency electric power generation: The environmental role," Prog Energy Combust Sci **33**(2), 107–134 (2007).
2. M.-H. Kim, B. Youn, and C.W. Bullard, "Effect of inclination on the air-side performance of a brazed aluminum heat exchanger under dry and wet conditions," International Journal of Heat and Mass Transfer **44**(24), 4613–4623 (2001)
3. N.V. Upot, K. Fazle Rabbi, S. Khodakarami, J.Y. Ho, J. Kohler Mendizabal, and N. Miljkovic, "Advances in micro and nanoengineered surfaces for enhancing boiling and condensation heat transfer: a review," Nanoscale Adv **5**(5), 1232–1270 (2022).
4. D. Xia, J. Yan, and S. Hou, "Fabrication of nanofluidic biochips with nanochannels for applications in DNA analysis," Small **8**(18), 2787–2801 (2012).
5. D. Vasileska, "Modeling thermal effects in nano-devices," Microelectron Eng **109**, 163–167 (2013).
6. B. Agostini, J.R. Thome, M. Fabbri, and B. Michel, "High heat flux two-phase cooling in silicon multimicrochannels," IEEE Transactions on Components and Packaging Technologies **31**(3), 691–701 (2008).
7. W.M.. Rohsenow, J.P.. Hartnett, and Y.I.. Cho, *Handbook of Heat Transfer* (McGraw-Hill, 1998).
8. A. Goswami, S.C. Pillai, and G. McGranaghan, "Surface modifications to enhance dropwise condensation," Surfaces and Interfaces **25**, (2021).
9. S. Qin, R. Ji, C. Miao, L. Jin, C. Yang, and X. Meng, "Review of enhancing boiling and condensation heat transfer: Surface modification," Renewable and Sustainable Energy Reviews **189**, (2024).
10. N. Miljkovic, R. Xiao, D.J. Preston, R. Enright, I. McKay, and E.N. Wang, "Condensation on hydrophilic, hydrophobic, nanostructured superhydrophobic and oil-infused surfaces," J Heat Transfer **135**(8), 1 (2013).
11. A. Starostin, V. Valtsifer, Z. Barkay, I. Legchenkova, V. Danchuk, and E. Bormashenko, "Drop-wise and film-wise water condensation processes occurring on metallic micro-scaled surfaces," Appl Surf Sci **444**, 604–609 (2018).
12. S. Daniel, M.K. Chaudhury, and J.C. Chen, "Fast drop movements resulting from the phase change on a gradient surface," Science (1979) **291**(5504), 633–636 (2001).
13. C. Dorrer, and J. Rühe, "Some thoughts on superhydrophobic wetting," Soft Matter **5**(1), 51–61 (2009).
14. S. Paul, D. Chakraborty, S.J. Esha, and M.N. Hasan, "Role of wettability contrast on nanoscale condensation over hybrid wetting surface with gradient and patterned wetting configuration at various philic-phobic content," Surfaces and Interfaces **36**, (2023).
15. L. Li, P. Ji, and Y. Zhang, "Molecular dynamics simulation of condensation on nanostructured surface in a confined space," Appl Phys A Mater Sci Process **122**(5), (2016).
16. A.K.M.M. Morshed, T.C. Paul, and J.A. Khan, "Effect of nanostructures on evaporation and explosive boiling of thin liquid films: A molecular dynamics study," Appl Phys A Mater Sci Process **105**(2), 445–451 (2011).
17. S. Ghahremanian, A. Abbassi, Z. Mansoori, and D. Toghraie, "Effect of nanostructured surface configuration on the interface properties and heat transfer of condensation process of argon inside nanochannels using molecular dynamics simulation," J Mol Liq **339**, 117281 (2021).
18. Z. Song, Z. Cui, Y. Liu, and Q. Cao, "Heat transfer and flow characteristics in nanochannels with complex surface topological morphology," Appl Therm Eng **201**, (2022).
19. A.P. Thompson, H.M. Aktulga, R. Berger, D.S. Bolintineanu, W.M. Brown, P.S. Crozier, P.J. in 't Veld, A. Kohlmeyer, S.G. Moore, T.D. Nguyen, R. Shan, M.J. Stevens, J. Tranchida, C. Trott, and S.J. Plimpton, "LAMMPS - a flexible simulation tool for particle-based materials modeling at the atomic, meso, and continuum scales," Comput Phys Commun **271**, (2022).
20. J.E. Lennard-Jones, and A.F. Devonshire, "Critical phenomena in gases - I," Proceedings of the Royal Society of London a Mathematical and Physical Sciences **163**(912), 53–70 (1937). "Critical phenomena in gases," Nature **141**(3582), 1148 (1938).
21. A. Hens, R. Agarwal, and G. Biswas, "Nanoscale study of boiling and evaporation in a liquid Ar film on a Pt heater using molecular dynamics simulation," Int J Heat Mass Transf **71**, 303–312 (2014).
22. A. Stukowski, "Visualization and analysis of atomistic simulation data with OVITO-the Open Visualization Tool," Model Simul Mat Sci Eng **18**(1), (2010).